\documentclass[a4paper,11pt]{article}
\usepackage[hmargin=1.5cm,vmargin=3cm]{geometry,}
\usepackage{graphicx, amssymb, textcomp,subfig, hyperref, amsmath}
\numberwithin{equation}{section}
\usepackage[titletoc]{appendix}
\title{\bf New quasi-exactly solvable class of generalized isotonic oscillators }
\author{\Large Davids Agboola,~Jon Links,~Ian Marquette,~Yao-Zhong Zhang}
\date{\it School of Mathematics and Physics, The University of Queensland, \\Brisbane, QLD 4072, Australia}
\begin{document}
\maketitle

\noindent{\bf Abstract}:~We introduce a new family of quasi-exactly solvable generalized isotonic oscillators which are based on the pseudo-Hermite exceptional orthogonal polynomials. We obtain exact closed-form expressions for the energies and wavefunctions as well as the allowed potential parameters for the first two members of the family using the Bethe ansatz method. 
Numerical calculations of the energies reveal that member potentials have multiple quasi-exactly solvable eigenstates and the number of states for higher members are parameter dependent.

\vspace{.3in}



\vskip.5in
\section{Introduction}
Due to their significance in aspects of exact solvability in quantum mechanics, exceptional orthogonal polynomials (EOPs) have attracted a wide range of interests \cite{CQ2008}-\cite{PTV2012}. Unlike the classical orthogonal polynomials, EOPs start with polynomials of degree one or higher. One essential characteristic of the EOPs is that they form complete orthogonal set with respect to some positive definite measure.
Recent attention has been on the construction of new one-dimensional quantum systems involving EOPs. Such constructions are most often discussed in connection with supersymmetric quantum mechanics \cite{CQ2008} and multi-dimensional superintegrable systems with higher-order integrals of motion \cite{MQ2013a,MQ2013c}. Moreover, it has been shown that exactly solvable quantum systems related to EOPs also allow different type of ladder operators with infinite sequences or multiplet states \cite{MQ2013b,MQ2013c}. In addition, some very interesting and unusual properties of superintegrable systems which relate to the EOPs have been discussed very recently \cite{MQ2013c}. 

An interesting feature of quantum mechanical models  based on certain EOPs is that the  corresponding Schr\"odinger equation can be reduced to differential equations which possess exact, partially, algebraically solvable spectra. Such systems are said to be Quasi-Exactly Solvable (QES). Thus a quantum mechanical system is said to be quasi-exactly solvable if only a finite number of eigenvalues and corresponding eigenvectors can be obtained exactly through algebraic means \cite{Turbiner96,GKO93,Ushveridze94}. A fundamental characteristic of QES systems is that the coefficients of the power series solutions to the underlying differential equations satisfy three-terms or more recursion relations, in contrast to the two-term recursions for exactly solvable cases. The complexity of higher order recursion relations makes it difficult to get power series solutions of such systems. In some cases, however, one can terminate the infinite series at certain power  by imposing certain constraints on the system parameters.  By so doing, analytic (polynomial) solutions to the systems can be obtained under certain constraints on the potential parameters.

However, the recently introduced functional Bethe ansatz method \cite{YZZ2012} (also see appendix for generalization) has proven very effective in obtaining exact closed-form polynomial solutions to many QES quantum mechanical model 
\cite{AZ11}-\cite{AZMPL12}. The aim of this paper is to obtain exact solutions of a class of quantum models based on the pseudo-Hermite EOP by means of the Bethe ansatz method \cite{YZZ2012}. 

The work is  organized as follows. In section 2, we transform the Schr\"odinger equation for a family of quantum systems into a QES differential equation. The general, closed form expressions for the energies, wavefunctions and the allowed parameters for the first two members of the family are derived in sections 3 and 4 respectively. We conclude the work in section 5, with summary and some remarks.

\section{The models and the underlying differential equation}
We consider a family of quantum systems with potentials
\begin{equation}\label{eq:1}
V_m(r)=\frac{1}{2}\left[\sum_{k=1}^{m-1}\mathcal{A}_kr^{2k}+g\left(\frac{\mathcal{H}_m''}{\mathcal{H}_m}-\left(\frac{\mathcal{H}_m''}{\mathcal{H}_m}\right)^2\right)\right],
\end{equation} where $m=2,4,6,\dots$ is an even integer, $r\in(0,\infty)$, $\mathcal{A}_k$ represent potential parameters, $g$ is a constant and $\mathcal{H}_m$ is the pseudo-Hermite polynomial of degree $m$ with all its coefficients positive, defined  by
\begin{equation}\label{eq:2}
\mathcal{H}_m(r)=m!\sum_{p=0}^{\frac{m}{2}}\frac{(2r)^{m-2p}}{p!(m-2p)!},~~m=2,4,6,\dots
.\end{equation} 
In recent papers \cite{CPRS08,FS2009,S2010}, cases of the first member potential ($m=2$) 
\begin{equation}\label{eq:3a}
V_2(x)=\frac{\omega^2x^2}{2}+g_a\frac{x^2-a^2}{(x^2+a^2)^2}
\end{equation}
where $a$ is a positive parameter have been studied. Particularly in \cite{CPRS08}, the model was shown to be exactly solvable in the case of $g_a=2$ and $\omega a^2=1/2$, with a general solution 
\begin{equation}\label{eq:3b}
\left\{\begin{array}{lll}
\Psi_n(x)=\frac{P_n(x)}{(2x^2+1)}e^{-x^2/2},\\\\
E_n=-\frac{3}{2}+n ,\hspace{0.2in} n=0,3,4,5,\dots\end{array}\right..
\end{equation}
 where $\Psi_n(x)$ is the wavefunction and $E_n$ is the energy and the polynomial $P_n(x)$ relates to the Hermite polynomial $H_n(x)$ as follows
\begin{equation}\label{eq:3c}
P_n(x)=\left\{\begin{array}{lll}
1\hspace{3in}\mbox{for}\hspace{0.1in}n=0,\\\\
H_n(x)+4nH_{n-2}(x)+4n(n-3)H_{n-4}(x)\hspace{0.35in}\mbox{for}\hspace{0.1in}n=3,4,5,\dots,\end{array}\right.
\end{equation}In what follows, we show that the isotonic oscillator is a member of a solvable class of EOP potentials \eqref{eq:1}, of which we shall obtain the solutions of the first two members using the Beth ansatz method.

The Schr\"odinger equation corresponding to the potential  Eq.\,\eqref{eq:1} is given by ($\hbar=\mu=1$)
\begin{equation}\label{eq:4}
\left[-\frac{d^2}{dr^2}+\frac{\ell(\ell+1)}{r^2}+\sum_{k=1}^{m-1}\mathcal{A}_kr^{2k}+g\left(\frac{\mathcal{H}_m''}{\mathcal{H}_m}-\left(\frac{\mathcal{H}_m''}{\mathcal{H}_m}\right)^2\right)\right]\Psi(r)=2E\Psi(r),
\end{equation} where $\ell=-1,0,1,\dots$, $E$ is the energy eigenvalue and $\Psi(r)$ is the wavefunction. After a brief inspection of the differential equation, we use the transformation 
\begin{equation}\label{eq:5}
\Psi(r)=r^{\ell+1}\mathcal{H}_m^\nu e^{w(r)}f(r),~~w(r)=\sum_{p=1}^{\frac{m}{2}}\mathcal{B}_pr^{2p},~~\nu=\frac{1}{2}\left[1-\sqrt{1-4g}\right],  ~m=2,4,\dots,
\end{equation}
to reduce  Eq.\,\eqref{eq:4} to 
\begin{equation}\label{eq:6}
f''(r)+2\left[\frac{\ell+1}{r}+\nu\frac{\mathcal{H}_m'}{\mathcal{H}_m}+w'(r)\right]f'(r)+\left[2\nu\left(w'(r)+\frac{\ell+1}{r}\right)\frac{\mathcal{H}_m'}{\mathcal{H}_m}+(\nu-g)\frac{\mathcal{H}_m''}{\mathcal{H}_m}+2w'(r)\frac{\ell+1}{r}\right.$$$$\left. \left(w'(r)^2+w''(r)\right)-\sum_{k=1}^{m-1}\mathcal{A}_kr^{2k}\right]f(r)=-2Ef(r),
\end{equation}where $\mathcal{B}_p$ are some constants related to $\mathcal{A}_k$. 
\section{The $\mathcal{H}_2$ model: Quantum isotonic oscillator}
For $m=2$, we have the pseudo-Hermite polynomial $\mathcal{H}_2=4r^2+2$, with the potential
\begin{equation}\label{eq:8}
V_2(r)=\frac{1}{2}\omega r^2-2g\frac{2r^2-1}{(2r^2+1)^2}.
\end{equation}
This is the quantum isotonic oscillator studied in \cite{AZ11,HSY10,SHCY11}. The corresponding equation \eqref{eq:6} for potential $V_2(r)$ reads 
\begin{equation}\label{eq:9}
f''(r)+2\left[\frac{\ell+1}{r}+\frac{4\nu r}{2r^2+1}+2\alpha r\right]f'(r)+4\left[\frac{2\nu(2\alpha r^2+\ell+1)+\nu-g}{2r^2+1}+\alpha\left(\ell+\frac{3}{2}\right)\right.$$$$\left.~~~~~~+\alpha^2r^2-\frac{\omega^2}{4}r^2\right]f(r)=-2Ef(r),
\end{equation} where by comparism with  Eq.\,\eqref{eq:6}, we have  $w(r)=\alpha r^2$.
Thus if we take $\alpha=-\frac{\sqrt{\omega}}{2}$ and introduce the new variable $z=r^2$, we have 
\begin{equation}\label{eq:10}
2z(2z+1)f''(z)+\left[-4\sqrt{\omega}z^2+(4\ell-2\sqrt{\omega}+8\nu+3)z+2\ell+3\right]f'(z)$$$$
+\left[\left(2E-\sqrt{\omega}(4\nu+2\ell+3)\right)z+4\nu(\ell+1)-\frac{\sqrt{\omega}}{2}\left(\ell+\frac{3}{2}\right)+\frac{E}{2}-2g\right]f(z)=0
\end{equation}
 Eq.\,\eqref{eq:10} is quasi-exactly solvable and therefore possesses polynomial solutions of degree $n\geq 0$, which we write in the form 
\begin{equation}\label{eq:11}
f(z)=\prod_{i=1}^n(z-z_i),~~~~f(z)\equiv 1~~\mbox{for}~~ n=0,
\end{equation} where $\{z_i\}$ are the roots of the polynomial to be determined. To solve  Eq.\,\eqref{eq:10}, we apply the functional Bethe ansatz method. Substituting \eqref{eq:11} into \eqref{eq:10}, we obtain the energies and wavefunction
\begin{equation}\label{eq:12}
E_n=\sqrt{\omega}\left(2n+\ell+\frac{5}{2}-\sqrt{1-4g}\right)=\sqrt{\omega}\left(2n+2\nu+\ell+\frac{3}{2}\right),$$$$
\Psi_n(r)\sim r^{\ell+1}(2r^2+1)^\nu e^{-\frac{\sqrt{\omega}}{2}r^2}\prod_{i=1}^n(r^2-z_i)
\end{equation}subject to the constraint
\begin{equation}\label{eq:13}
\nu^2+\nu\left(2\ell+\frac{\sqrt{\omega}}{2}+1\right)=2\sqrt{\omega}\sum_{i=1}^nz_i-n\left(2n+2\ell+4\nu-\frac{\sqrt{\omega}}{2}-\frac{1}{2}\right)
\end{equation}with $\{z_i\}$ satisfying the equations
\begin{equation}\label{eq:14}
\sum_{j\neq i}^n\frac{2}{z_i-z_j}=\frac{4\sqrt{\omega}z_i^2-(4\ell-2\sqrt{\omega}+8\nu+3)z_i-2\ell-3}{2z_i(2z_i+1)}, ~~i=1,2,\dots,n.
\end{equation}
As examples, we shall obtain the solutions for $n=0,1,2$. For $n=0$, we have
\begin{equation}\label{eq:15}
E_0=\sqrt{\omega}\left(2\nu+\ell+\frac{3}{2}\right),$$$$
\Psi_0(r)\sim r^{\ell+1}(2r^2+1)^\nu e^{-\frac{\sqrt{\omega}}{2}r^2}
\end{equation} subject to the constraint
\begin{equation}\label{eq:16}
\nu=0~(g=0) ~~~\mbox{or}~~~\nu=-\left(2\ell+\frac{\sqrt{\omega}}{2}+1\right).
\end{equation} Similarly for $n=1$, we have the energy and wavefunction
\begin{equation}\label{eq:17}
E_1=\sqrt{\omega}\left(2\nu+\ell+\frac{7}{2}\right)$$$$
\Psi_1(r)\sim r^{\ell+1}(2r^2+1)^\nu e^{-\frac{\sqrt{\omega}}{2}r^2}(r^2-z_1)
\end{equation}subject to the constriant
\begin{equation}\label{eq:18}
\nu^2+\nu\left(2\ell+\frac{\sqrt{\omega}}{2}+5\right)=2\sqrt{\omega}z_1-\left(2\ell-\frac{\sqrt{\omega}}{2}+\frac{3}{2}\right)
\end{equation} where $z_1$ satisfies 
\begin{equation}\label{eq:19}
4\sqrt{\omega}z_1^2-(4\ell-2\sqrt{\omega}+8\nu+3)z_1-2\ell-3=0.
\end{equation}Equations \eqref{eq:18} and \eqref{eq:19} give the condition 
\begin{equation}\label{eq:20}
\nu^2+\nu\left(2\ell+\frac{\sqrt{\omega}}{2}+3\right)=\left(-\ell-\frac{3}{4}\pm\frac{1}{4}\sqrt{(4\ell-2\sqrt{\omega}+8\nu+3)^2+16(2\ell+3)\sqrt{\omega}}\right).
\end{equation}

\begin{table}[h]
\centering
\begin{tabular}{c c c c c c c c c c}\hline
$ {\ell}$&\hfil$\nu^{--}$ &\hfil ${E}_{1,\ell}^{--}$& \hfil $\nu^{-+}$&\hfil ${E}_{1,\ell}^{-+}$& \hfil $\nu^{+-}$&\hfil ${E}_{1,\ell}^{+-}$& \hfil $\nu^{++}$&\hfil ${E}_{1,\ell}^{++}$\\\hline
1&-3.04984&-0.50587&0.07658&1.47146&-0.676050&0.99545&-6.66690&-2.79351\\
2&-5.09883&-1.48553&0.04369&1.76688&-0.719447&1.28423&-8.54164&-3.66295\\
3&-7.11723&-2.44585&0.02961&2.07420&-0.764277&1.57211&-10.4643&-4.56274\\
4&-9.12691&-3.40065&0.02211&2.38569&-0.799033&1.86636&-12.4124&-5.47858\\
5&-11.1329&-4.35311&0.01754&2.69903&-0.825598&2.16578&-14.3753&-6.40380\\
6&-13.1369&-5.30436&0.01449&3.01333&-0.846266&2.46894&-16.3475&-7.33492\\
7&-15.1399&-6.25490&0.01233&3.32819&-0.86270&2.77477&-18.3260&-8.26998\\
8&-17.1421&-7.20499&0.01071&3.64339&-0.876036&3.08257&-20.3088&-9.20780\\
9&-19.1438&-8.15477&0.00946&3.95883&-0.887055&3.39182&-22.2948&-10.1476\\
10&-21.1452&-9.10434&0.00847&4.27443&-0.896301&3.70220&-24.2832&-11.0890\\\hline

\end{tabular}
\caption{$E_{1,\ell}$ levels for $\mathcal{H}_2$ model with $\omega=0.1$. The ''+'' and ''-'' superscripts represent the possible roots of  Eq.\,\eqref{eq:20}. }

\label{tab:tab1}
\end{table}

\begin{figure}[h]
\centering
\includegraphics[scale=.6]{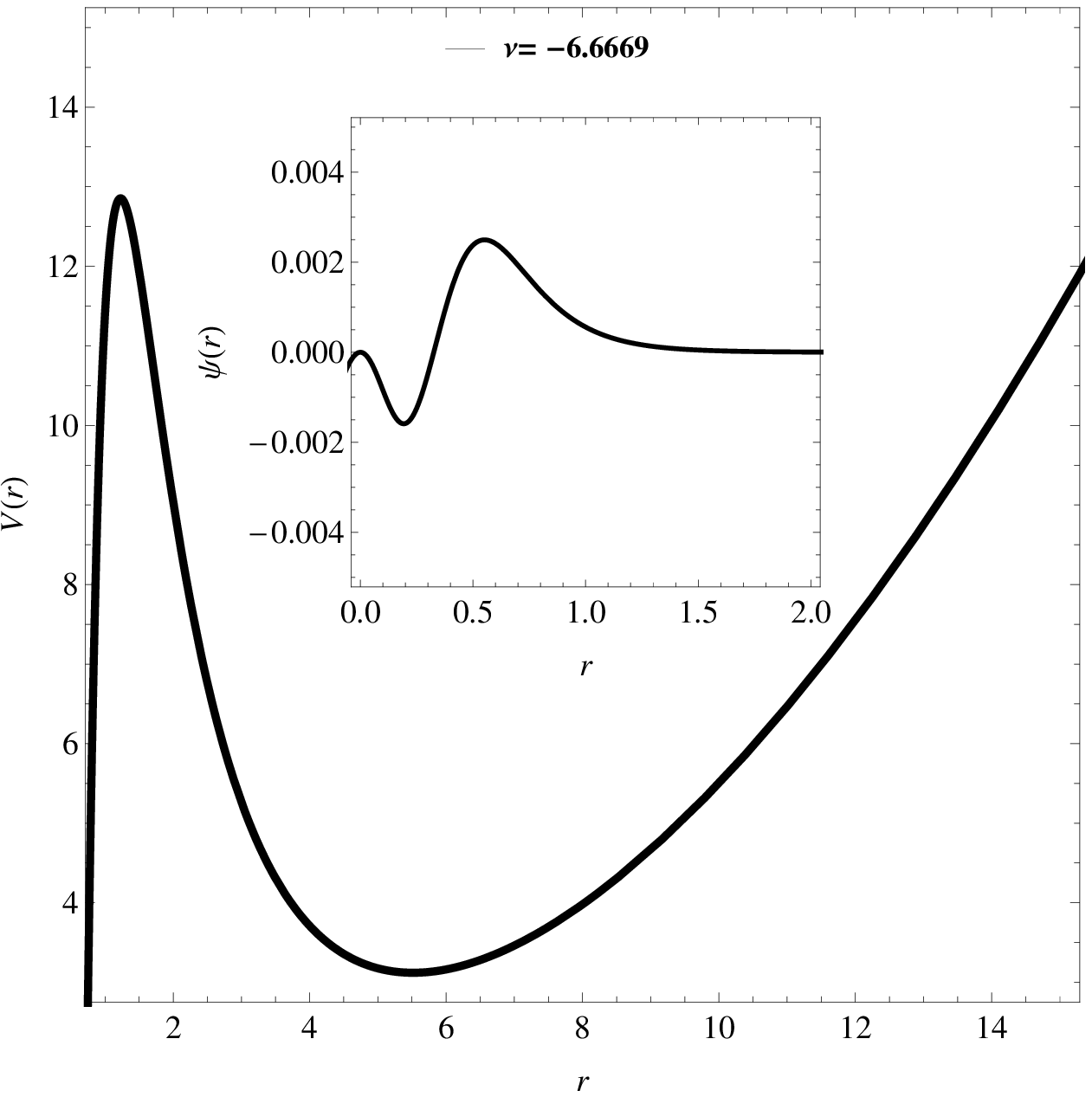}
\includegraphics[scale=.6]{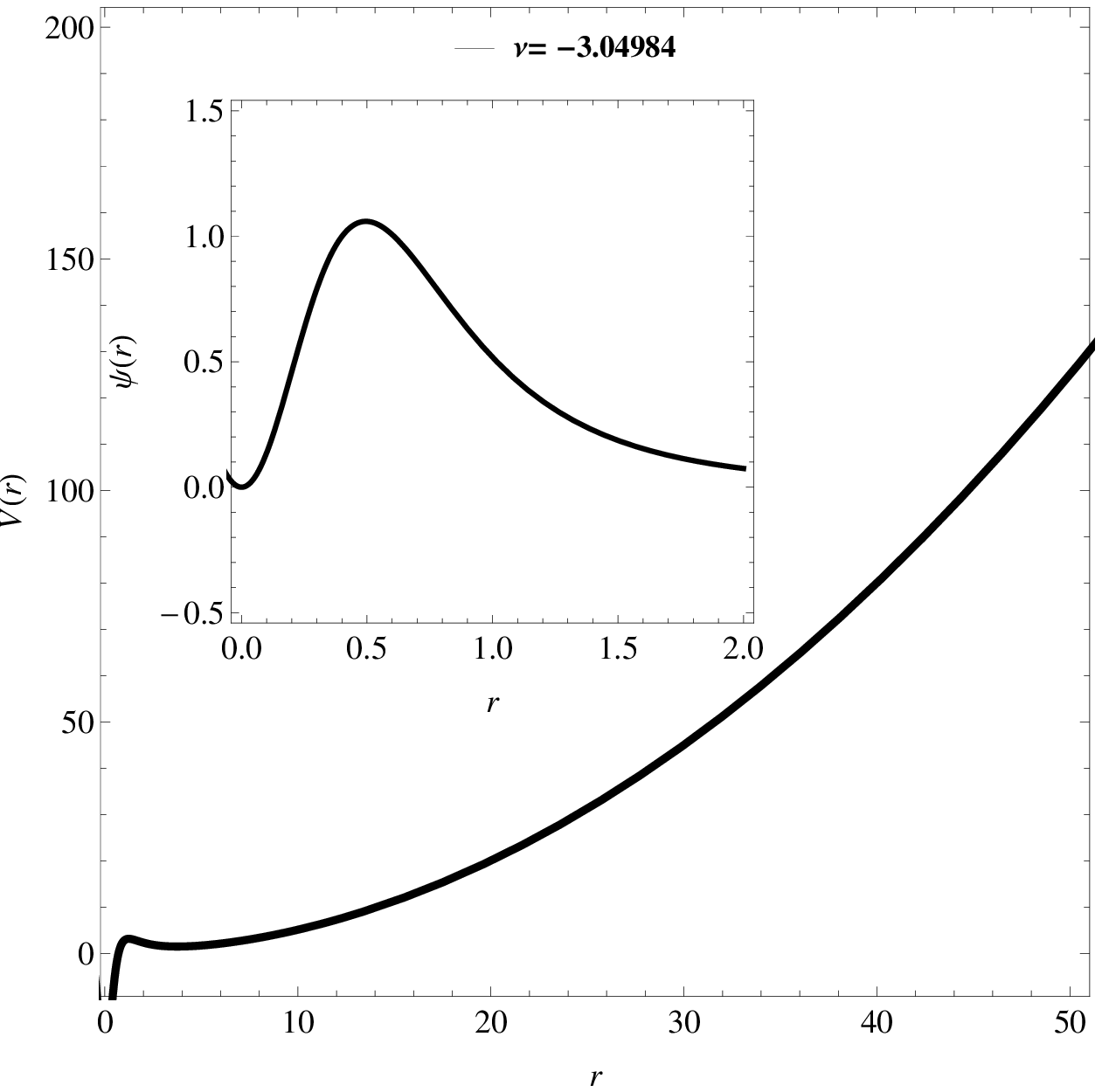}\\
\includegraphics[scale=.6]{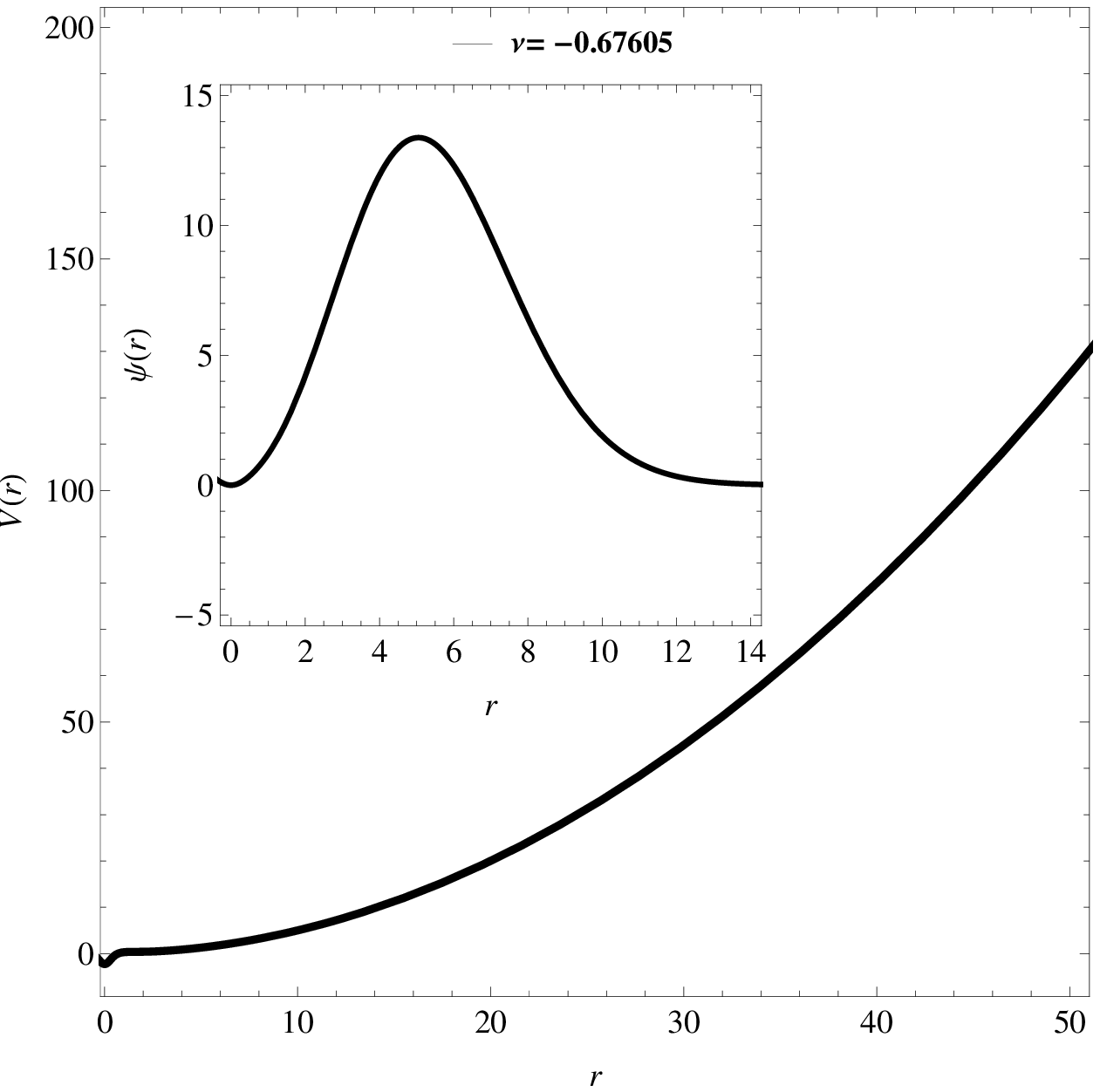}
\includegraphics[scale=.6]{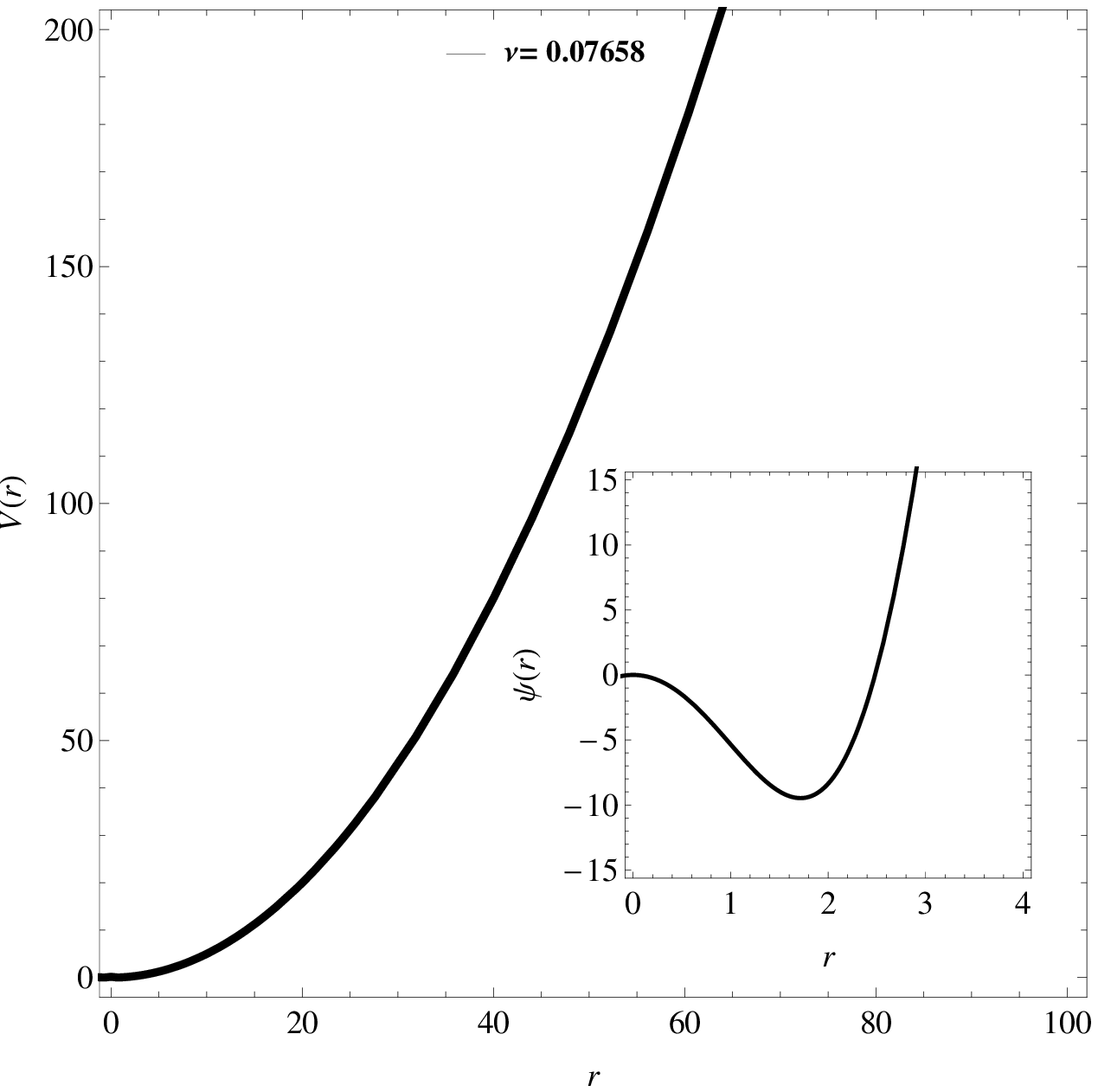}
\caption{Potential plots and corresponding wavefunctions (inserts) for $\mathcal{H}_2$ model with ($\omega$, $n$, $\ell$)=(0.1, 1, 1). }
\label{fig:}
\end{figure}

\noindent Also, for $n=2$, we have the solutions
\begin{equation}\label{eq:21}
E_2=\sqrt{\omega}\left(2\nu+\ell+\frac{9}{2}\right)$$$$
\Psi_2(r)\sim r^{\ell+1}(2r^2+1)^\nu e^{-\frac{\sqrt{\omega}}{2}r^2}(r^2-z_1)(r^2-z_2)
\end{equation}with the constraint
\begin{equation}\label{eq:22}
\nu^2+\nu\left(2\ell+\frac{\sqrt{\omega}}{2}+9\right)=2\sqrt{\omega}(z_1+z_2)-\left(2\ell-\frac{\sqrt{\omega}}{2}+\frac{7}{2}\right)
\end{equation}\\\\\\ where $z_1,z_2$ satisfy
\begin{equation}\label{eq:23}
\frac{2}{z_1-z_2}=\frac{4\sqrt{\omega}z_1^2-(4\ell-2\sqrt{\omega}+8\nu+3)z_1-2\ell-3}{2z_1(2z_1+1)},$$$$
\frac{2}{z_2-z_1}=\frac{4\sqrt{\omega}z_2^2-(4\ell-2\sqrt{\omega}+8\nu+3)z_2-2\ell-3}{2z_2(2z_2+1)}.
\end{equation}We now note the following: for physically valid solutions, $\nu\leq \frac{1}{2}$ and for numerical evaluation of the energies $E_0$ and $E_1$, $\omega$ can take any value. However, for the second excited state, one must carefully select the value of $\omega$ as this determines the validity of the solution. Thus in tables 1 and 2, by carefully selecting $\omega=0.1$, we give some numerical values of the energies and the allowed potential parameter for $\ell=0\dots 10$, for $n=1$ and 2 respectively. Moreover as shown in Fig 1, the potential well becomes shallower as the parameter $\nu$ (or $g$) increases. 


\begin{table}[htt]

\centering
\begin{tabular}{c c c c c c c}\hline
$ {\ell}$&$\nu_a$&$E_a$&$\nu_b$& $E_b$&$\nu_c$&$E_c$\\\hline
1&-10.6691&-5.00847&-7.39674&-2.93886&0.22493&1.88151\\
2&-12.5857&-5.90443&-9.34834&-3.85693&0.18139&2.1702\\
3&-14.5252&-6.81480&-11.3161&-4.78522&0.15198&2.46783\\
4&-16.4793&-7.73449&-13.2931&-5.71938&0.13073&2.77062\\
5&-18.4435&-8.66052&-15.2760&-6.65721&0.11466&3.07668\\
6&-20.4148&-9.59104&-17.2627&-7.59747&0.10208&3.38495\\
7&-22.3912&-10.5248&-19.2520&-8.53944&0.09198&3.69479\\
8&-24.3716&-11.4611&-21.2434&-9.48264&0.08368&4.00577\\
9&-26.3550&-12.3993&-23.2362&-10.4268&0.07675&4.31762\\
10&-28.3408&-13.3390&-25.2301&-11.3716&0.07088&4.63013\\\hline

\end{tabular}
\caption{Non-degenerate $E_{2,\ell}$ levels for $\mathcal{H}_2$ model with $\omega=0.1$}

\label{tab:tab3}
\end{table}
\section{The $\mathcal{H}_4$ Model: Deformed isotonic oscillator}
Similarly, for $m=4$ we have the pseudo-Hermite polynomial $\mathcal{H}_4=16r^4+48r^2+12$, such that the corresponding potential reads
\begin{equation}\label{eq:24}
V_4(r)=\frac{1}{2}\left[\gamma r^6+\rho r^4+\kappa r^2+g\frac{(96r^6+336r^4-2r^3+216r^2-3r+36)}{2(4r^4+12r^2+3)^2}\right],
\end{equation}with the corresponding equation
\begin{equation}\label{eq:25}
f''(r)+2\left[\frac{\ell+1}{r}+\nu\frac{48r^2+24}{4r^4+12r^2+3}+4\delta r^3+2\beta r\right]f'(r)+16\left[\frac{\nu(2r^2+3)(4\delta r^4+2\beta r^2+\ell+1)+(3r^2+2)(\nu-g)}{4r^4+12r^2+3}\right.$$$$\left.+4(\beta+2\gamma r^2)(\ell+1)+(4\delta r^3+2\beta r)^2+2\beta+12\delta r-\gamma r^6-\rho r^4-\kappa r^2\right]f(r)=-2Ef(r).
\end{equation} By comparism with  Eq.\,\eqref{eq:6}, we have $w(r)=\delta r^4+\beta r^2$. Thus, if we choose $\delta=-\frac{\sqrt{\gamma}}{4}$ and $\beta=-\frac{\rho}{4\sqrt{\gamma}}$, and introduce the new variable $z=r^2$, we have 
\begin{equation}\label{eq:26}
(4z^3+12z^2+3z)f''(z)+\left[-4\sqrt{\gamma}z^4-\left(12\sqrt{\gamma}+\frac{2\rho}{\sqrt{\gamma}}\right)z^3+\left(6-3\sqrt{\gamma}+4\ell+16\nu-\frac{6\rho}{\sqrt{\gamma}}\right)z^2\right.$$$$\left.+3\left(6+4\ell+8\nu-\frac{\rho}{2\sqrt{\gamma}}\right)z+3\left(\ell+\frac{3}{2}\right)\right]f'(z)+\left[\left(\frac{\rho^2}{4\gamma}-\sqrt{\gamma}(5+2\ell+8\nu)-\kappa\right)z^3\right.$$$$\left.\left(2E-3\sqrt{\gamma}(2\ell+4\nu+5)-\frac{\rho}{\sqrt{\gamma}}\left(\ell+4\nu+\frac{3}{2}\right)-\frac{3\rho^2}{4\gamma}-3\kappa\right)z^2\right.$$$$+\left.\left(6E-12g+\left(4\nu-\frac{3\sqrt{\gamma}}{4}\right)(2\ell+5)-\frac{3\rho}{4\sqrt{\gamma}}\left(6+4\ell+8\nu+\frac{3\rho^2}{4}\right)-\frac{3\kappa}{4}\right)z\right.$$$$\left.+\frac{3E}{2}+6\nu(2\ell+3)-6g-\frac{3\rho}{4\sqrt{\gamma}}\left(\ell+\frac{3}{2}\right)\right]f(z)=0.
\end{equation}

\begin{figure}[h]
\centering
\includegraphics[scale=.6]{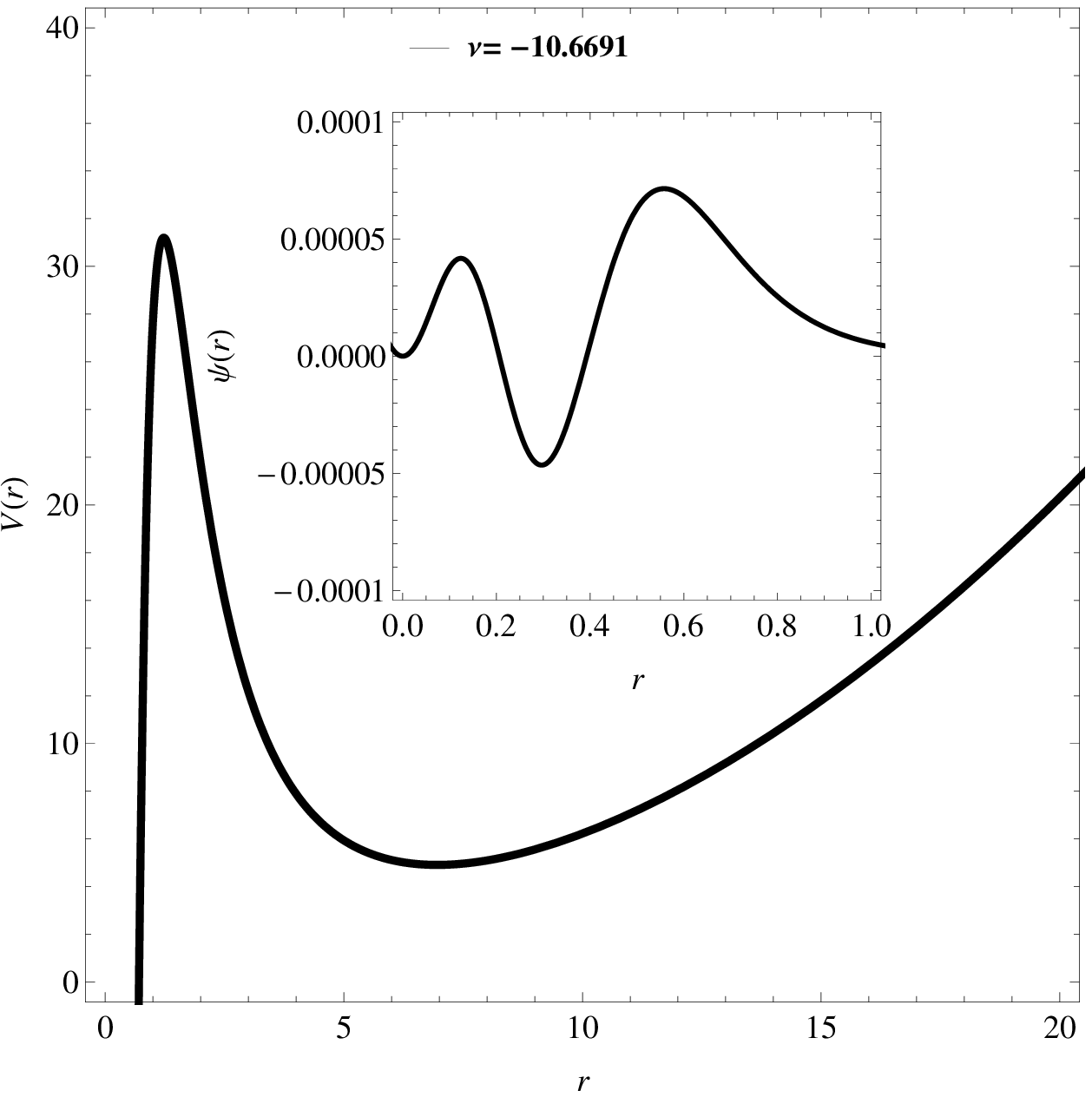}
\includegraphics[scale=.6]{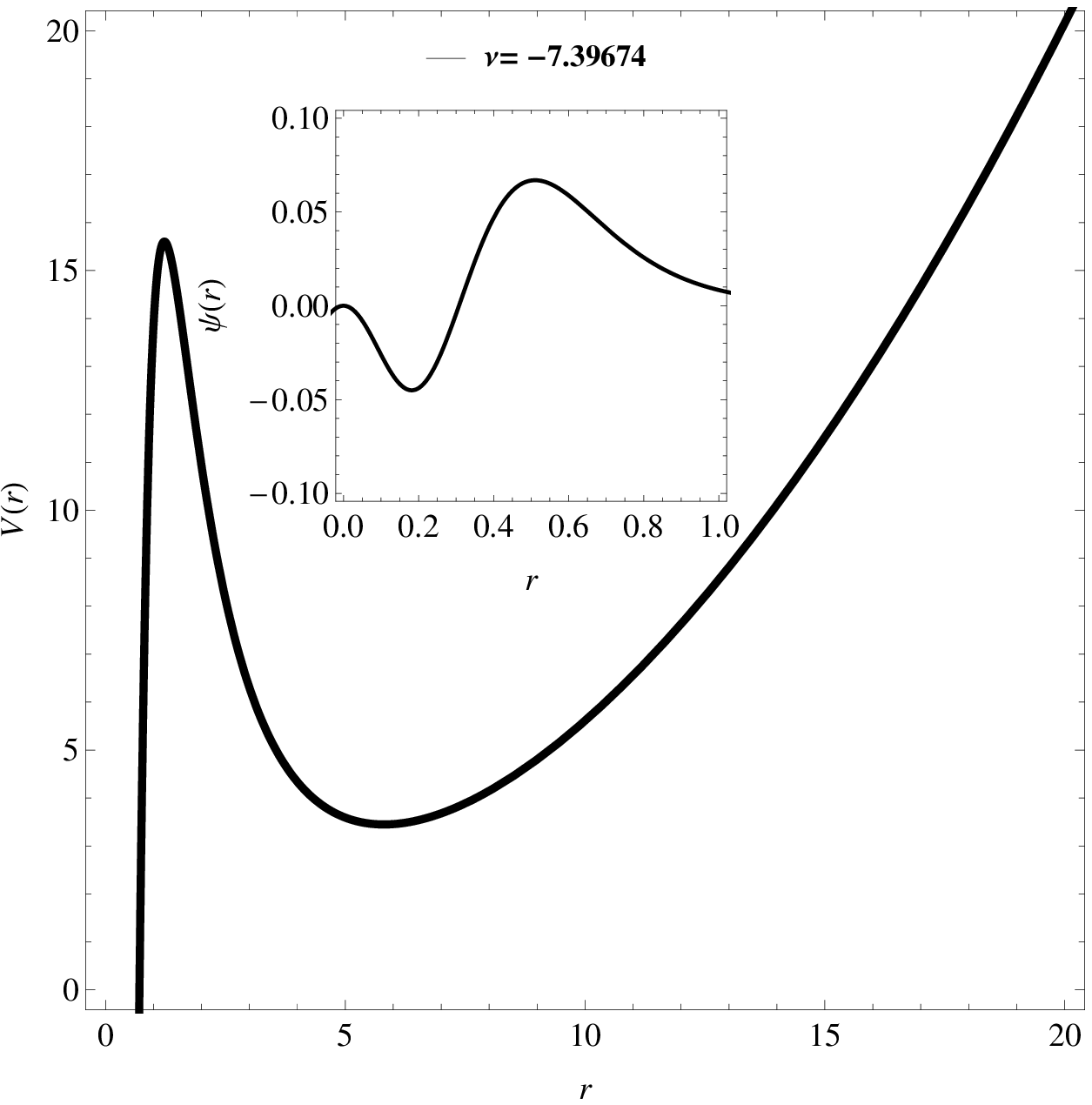}\\
\includegraphics[scale=.6]{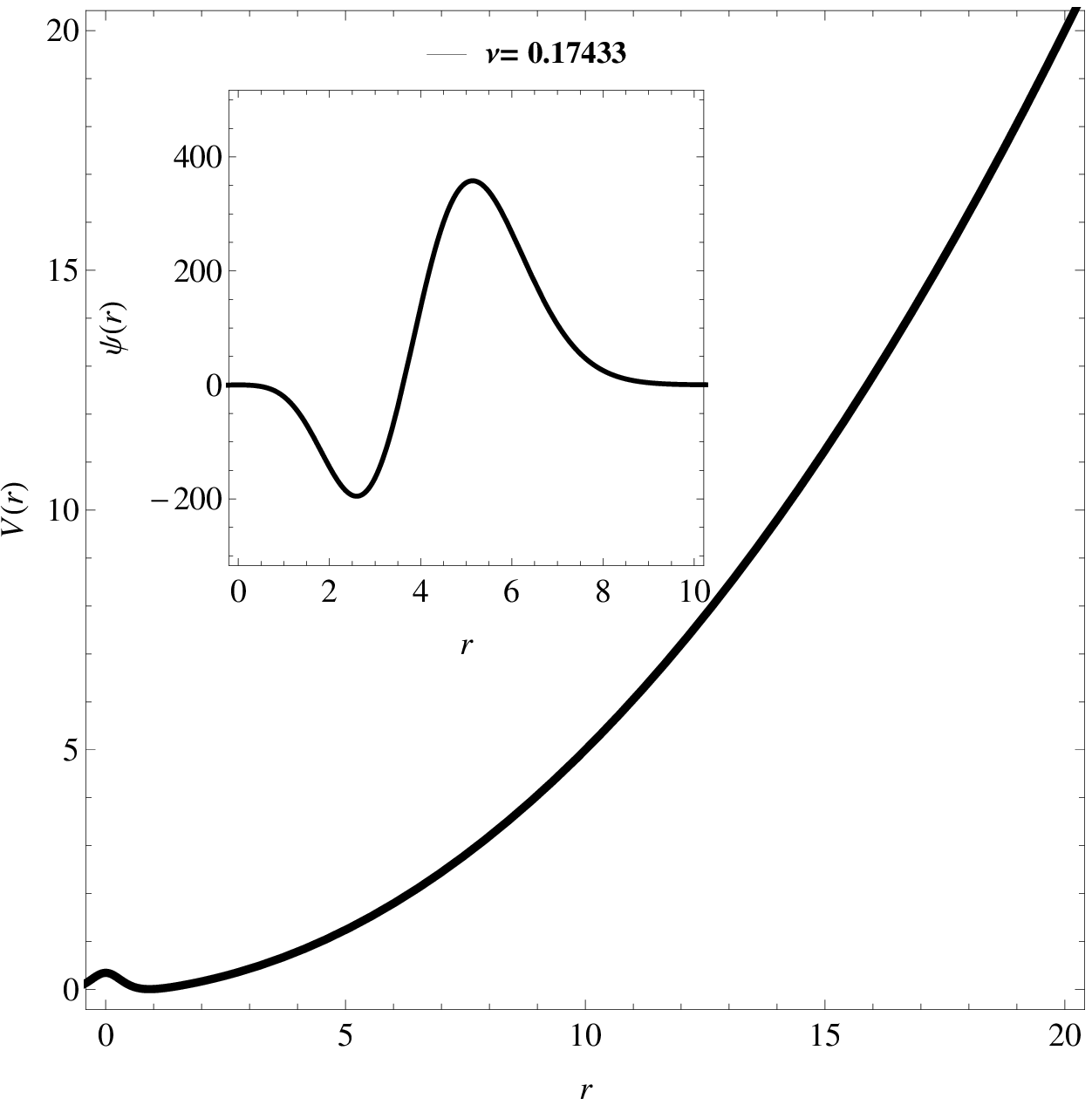}
\caption{\small Potential plots and corresponding wavefunctions (inserts) for $\mathcal{H}_2$ model with ($\omega$, $n$, $\ell$)=(0.1, 2, 1)}
\label{fig:}
\end{figure}

\noindent If we seek the solution of the form \eqref{eq:11}, we have the energy and the wavefunction
\begin{equation}\label{eq:27}
E_n=2\sqrt{\gamma}\sum_{i=1}^nz_i+3\sqrt{\gamma}\left(2n+2\nu+\ell+\frac{5}{2}\right)+\frac{\rho}{\sqrt{\gamma}}\left(n+2\nu+\frac{\ell}{2}+\frac{3}{4}\right)+\frac{3\rho^2}{8\gamma}+\frac{3\kappa}{2}$$$$
\Psi_n(r)\sim r^{\ell+1}(4r^4+12r^2+3)^\nu e^{-\frac{\sqrt{\gamma}}{4}r^4-\frac{\rho}{4\sqrt{\gamma}}r^2}\prod_{i=1}^n(r^2-z_i)
\end{equation}subject to the constriants
\begin{equation}\label{eq:28}
\kappa=\frac{\rho^2}{4\gamma}-\sqrt{\gamma}(4n+2\ell+8\nu+5),$$$$
-2\sqrt{\gamma}\sum_{i=1}^nz_i^3-\left(6\sqrt{\gamma}+\frac{\rho}{\sqrt{\gamma}}\right)\sum_{i=1}^nz_i^2+\left(4n+2\ell+8\nu-\frac{3\rho}{\sqrt{\gamma}}-\frac{3\sqrt{\gamma}}{2}-1\right)\sum_{i=1}^nz_i$$$$
+n\left(6\ell+12\nu-\frac{3\rho}{4\sqrt{\gamma}}+15\right)+\frac{3E_n}{4}+3\nu(2\ell+\nu+2)=0,$$$$
6E_n-12\nu(1-\nu)-\frac{3\kappa}{4}=4\sqrt{\gamma}\sum_{i=1}^nz_i^2+\left(6\sqrt{\gamma}+\frac{\rho}{\sqrt{\gamma}}\right)\sum_{i=1}^nz_i-n(4n+4\ell+16\nu-3\sqrt{\gamma}+2)$$$$
+\frac{\rho}{\sqrt{\gamma}}\left(6n+6\nu+2\ell-\frac{3\rho^2}{16}\right)-(2\ell+5)\left(4\nu-\frac{3\sqrt{\gamma}}{4}\right),
\end{equation}
with the roots $\{z_i\}$ satisfying the equations
\begin{equation}\label{eq:29}\small
\sum_{j\neq i}^n\frac{2}{z_i-z_j}=\frac{-4\sqrt{\gamma}z^4_i-\left(12\sqrt{\gamma}+\frac{2\rho}{\sqrt{\gamma}}\right)z^3_i+\left(6-3\sqrt{\gamma}+4\ell+16\nu-\frac{6\rho}{\sqrt{\gamma}}\right)z^2_i+3\left(6+4\ell+8\nu-\frac{\rho}{2\sqrt{\gamma}}\right)z_i+3\left(\ell+\frac{3}{2}\right)}{z_i(4z_i^2+12z_i+3)},~i=1,2,\dots,n.
\end{equation}

\noindent We now obtain the solutions corresponding to $n=0,1,2$. For $n=0$, we have the solutions
\begin{equation}\label{eq:30}
E_0=3\sqrt{\gamma}\left(2\nu+\ell+\frac{5}{2}\right)+\frac{\rho}{\sqrt{\gamma}}\left(2\nu+\frac{\ell}{2}+\frac{3}{4}\right)+\frac{3\rho^2}{8\gamma}+\frac{3\kappa}{2}$$$$
\Psi_0(r)\sim r^{\ell+1}(4r^4+12r^2+3)^\nu e^{-\frac{\sqrt{\gamma}}{4}r^4-\frac{\rho}{4\sqrt{\gamma}}r^2}
\end{equation}subject to the constriants
\begin{equation}\label{eq:31}
\kappa=\frac{\rho^2}{4\gamma}-\sqrt{\gamma}(2\ell+8\nu+5),$$$$
\frac{3E_0}{4}+3\nu(2\ell+\nu+2)=0,$$$$
6E_0-12\nu(1-\nu)-\frac{3\kappa}{4}=\frac{\rho}{\sqrt{\gamma}}\left(6\nu+2\ell-\frac{3\rho^2}{16}\right)-(2\ell+5)\left(4\nu-\frac{3\sqrt{\gamma}}{4}\right).
\end{equation}
Similar to the $\mathcal{H}_2$ model, the $n=0$ solution has two levels  for each $\ell$. The first level corresponds to $\nu=0$, while the numerical values for the second level are given in table 3.  
\begin{table}[htt]

\centering
\begin{tabular}{c c c c c}\hline
$ {\ell}$&\hfil$\rho$&\hfil $\kappa$&\hfil $\nu$&\hfil $E_{0,\ell}$\\\hline
1&-0.40708&-1.87883&0.03143&-0.50683\\
2&-0.46411&-2.43204&0.04921&-1.19070\\
3&-0.51625&-2.9686&0.06182&-1.99346\\
4&-0.56659&-3.49153&0.07238&-2.91624\\
5&-0.61608&-4.00181&0.08194&-3.95976\\
6&-0.66514&-4.49984&0.09092&-5.12448\\
7&-0.71397&-4.98579&0.09955&-6.41077\\
8&-0.76267&-5.45973&0.10795&-7.81894\\
9&-0.81130&-5.92166&0.11619&-9.34926\\
10&-0.85991&-6.37158&0.12432&-11.0020\\\hline

\end{tabular}
\caption{$E_{0,\ell}$ levels and allowed parameters $\nu$ and $\rho$ for $\mathcal{H}_4$ model
for $\gamma=0.1$.}

\label{tab:tab3}
\end{table}

\begin{figure}[ht]
\centering
\includegraphics[scale=.6]{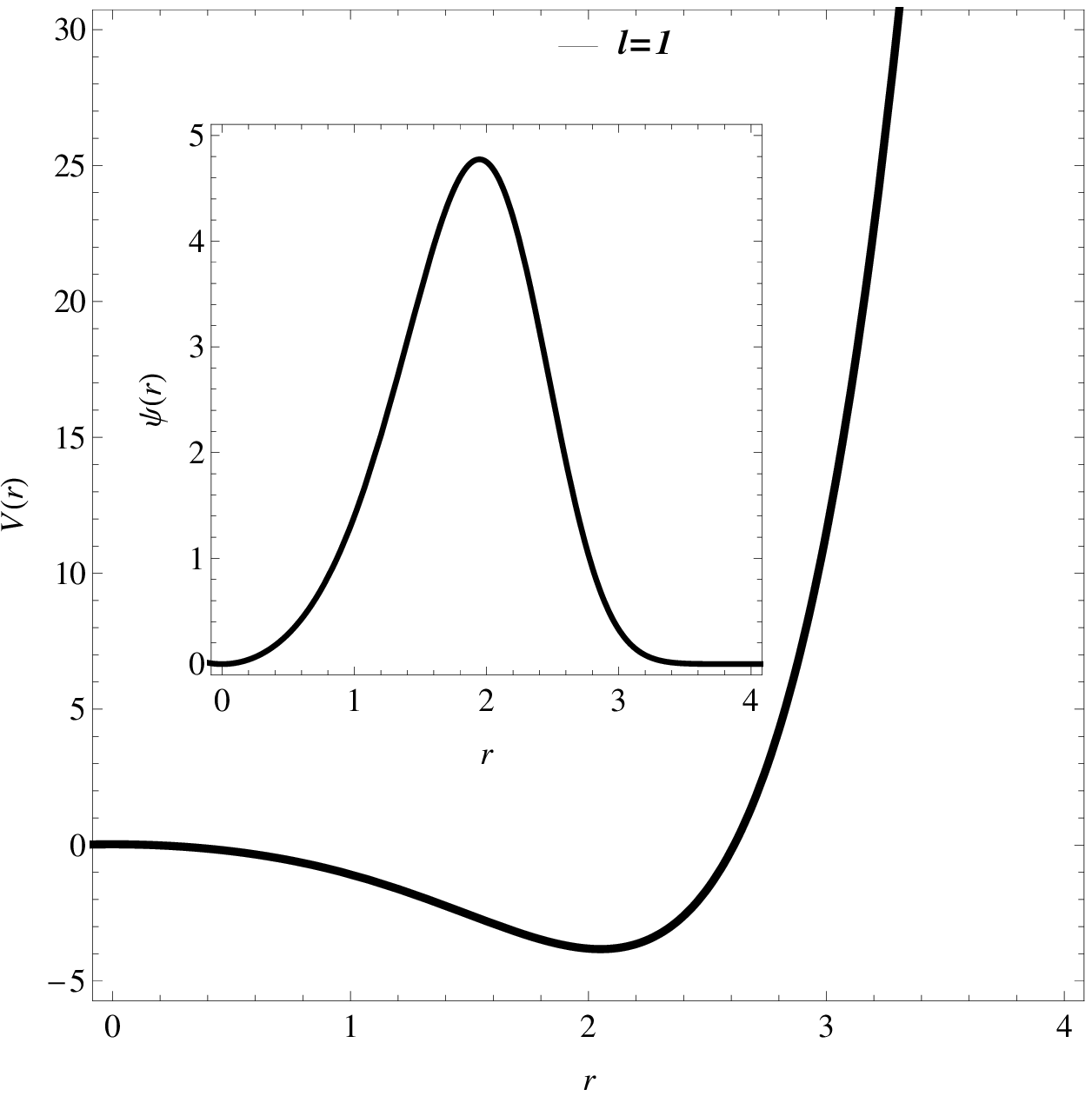}
\includegraphics[scale=.6]{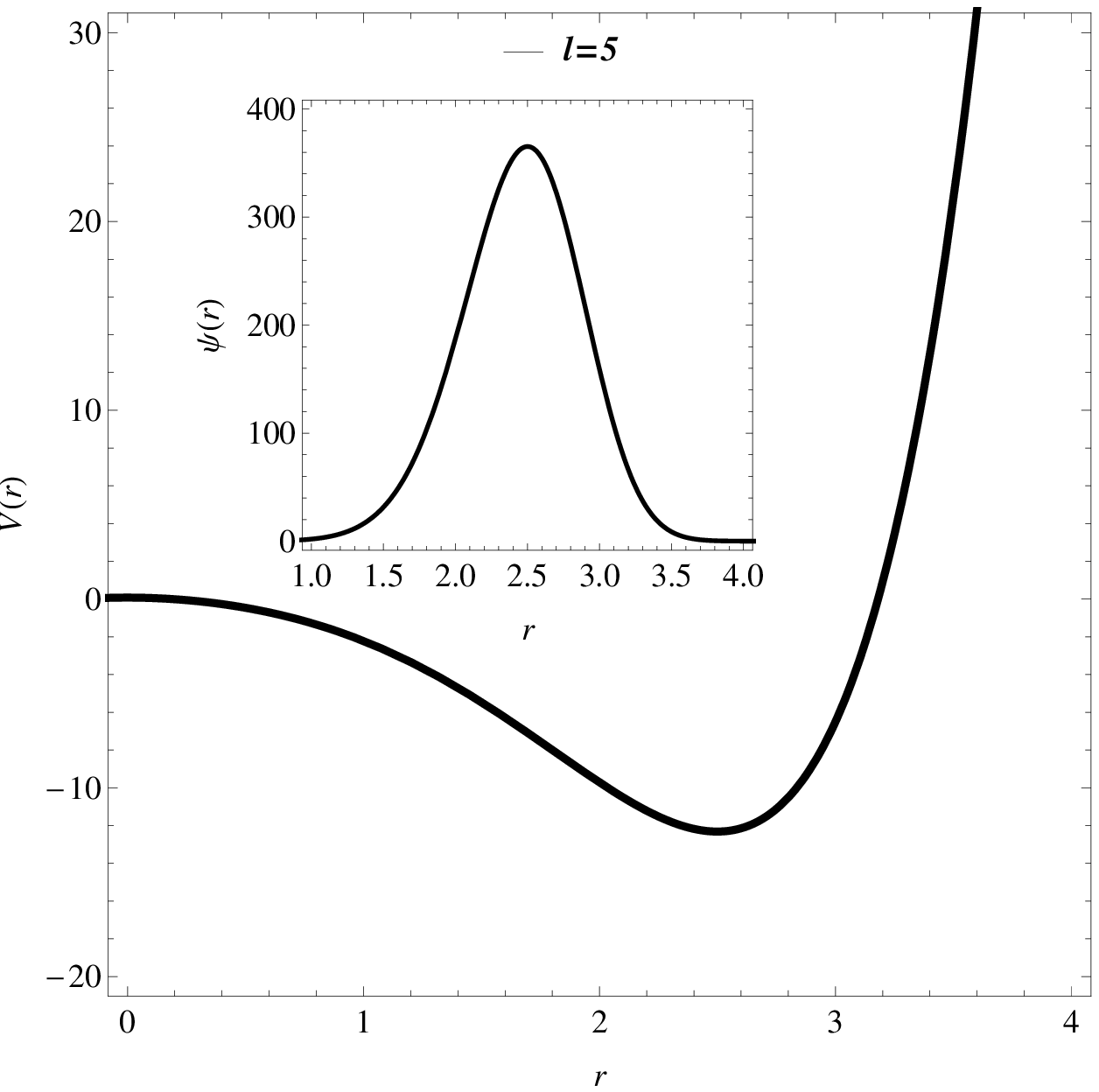}
\includegraphics[scale=.6]{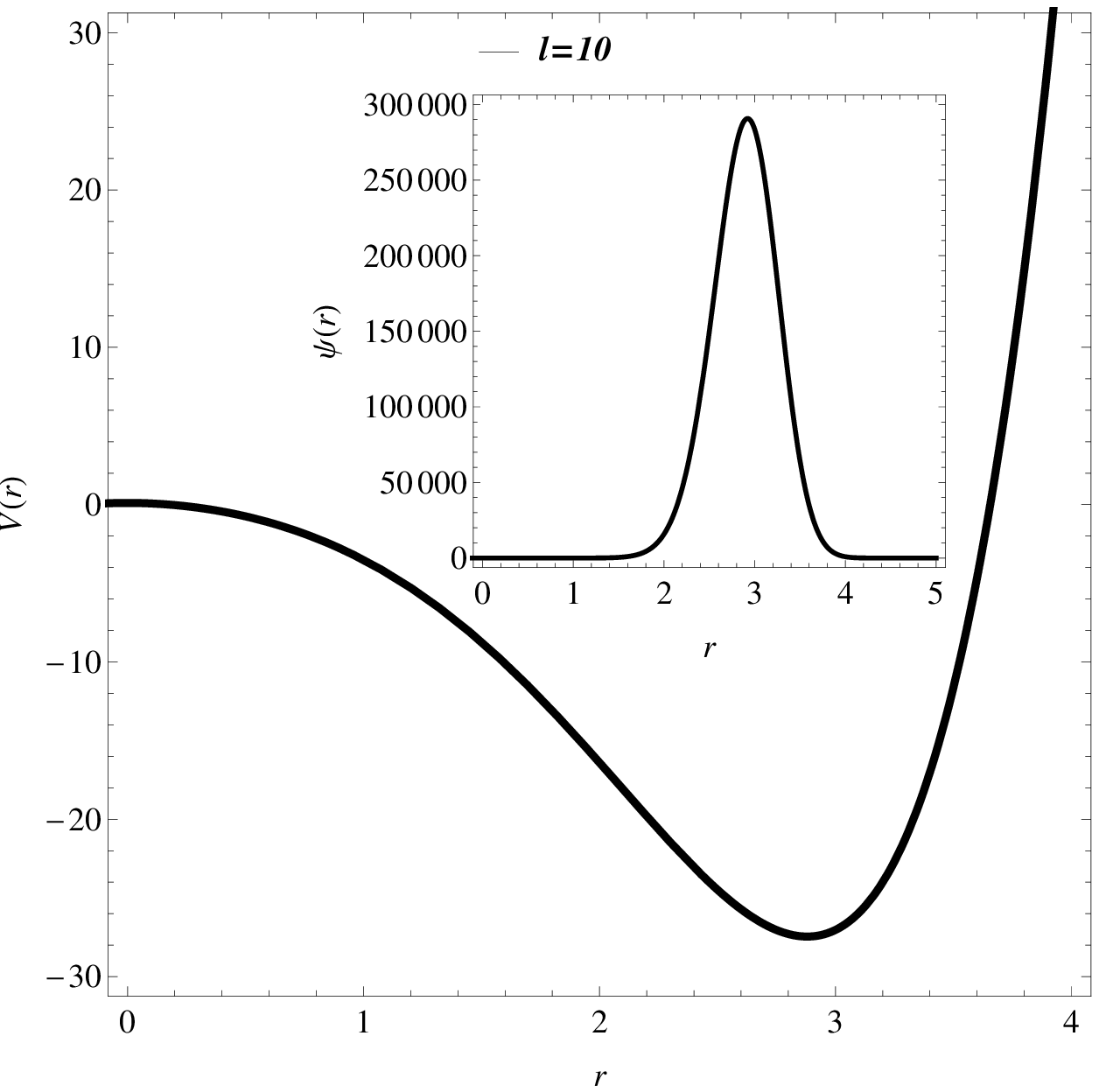}
\caption{\small Potential plots and corresponding wavefunctions (inserts) for $\mathcal{H}_4$ model with($\gamma$, $n$, $\ell$)=(0.1, 0) and $\ell=1, 5, 10$}
\label{fig:}
\end{figure}




Similarly for $n=1$, we have the solutions
\begin{equation}\label{eq:32}
E_1=2\sqrt{\gamma}z_1+3\sqrt{\gamma}\left(2\nu+\ell+\frac{9}{2}\right)+\frac{\rho}{\sqrt{\gamma}}\left(2\nu+\frac{\ell}{2}+\frac{7}{4}\right)+\frac{3\rho^2}{8\gamma}+\frac{3\kappa}{2}$$$$
\Psi_1(r)\sim r^{\ell+1}(4r^4+12r^2+3)^\nu e^{-\frac{\sqrt{\gamma}}{4}r^4-\frac{\rho}{4\sqrt{\gamma}}r^2}(r^2-z_1)
\end{equation}subject to the constriants
\begin{equation}\label{eq:33}
\kappa=\frac{\rho^2}{4\gamma}-\sqrt{\gamma}(2\ell+8\nu+9)$$$$
-2\sqrt{\gamma}z_1^3-\left(6\sqrt{\gamma}+\frac{\rho}{\sqrt{\gamma}}\right)z_1^2+\left(2\ell+8\nu-\frac{3\rho}{\sqrt{\gamma}}-\frac{3\sqrt{\gamma}}{2}+3\right)z_1$$$$
+\left(6\ell+12\nu-\frac{3\rho}{4\sqrt{\gamma}}+15\right)+\frac{3E_1}{4}+3\nu(2\ell+\nu+2)=0,$$$$
6E_1-12\nu(1-\nu)-\frac{3\kappa}{4}=4\sqrt{\gamma}z_1^2+\left(6\sqrt{\gamma}+\frac{\rho}{\sqrt{\gamma}}\right) z_1-(4\ell+16\nu-3\sqrt{\gamma}+6)$$$$
+\frac{\rho}{\sqrt{\gamma}}\left(6+6\nu+2\ell-\frac{3\rho^2}{16}\right)-(2\ell+5)\left(4\nu-\frac{3\sqrt{\gamma}}{4}\right),
\end{equation}
with the roots $z_1$ satisfying the equation
\begin{equation}\label{eq:34}\small
-4\sqrt{\gamma}z^4_1-\left(12\sqrt{\gamma}+\frac{2\rho}{\sqrt{\gamma}}\right)z^3_1+\left(6-3\sqrt{\gamma}+4\ell+16\nu-\frac{6\rho}{\sqrt{\gamma}}\right)z^2_i+3\left(6+4\ell+8\nu-\frac{\rho}{2\sqrt{\gamma}}\right)z_1+3\left(\ell+\frac{3}{2}\right)=0
\end{equation}

By solving equations \eqref{eq:33} and \eqref{eq:34} simultaneously using the numerical {\tt Mathematica} function {\tt NSolve}, which is program to give a more complete set of solutions for multivariate nonlinear algebraic equations, we obtain the allowed values of the potential parameters. We observe that as we change the parameter $\gamma$ from 1 to 100, the total number of solutions (real and complex) changes while a change in the angular momentum $\ell$ alters the number of real solutions for any given $\gamma$. By ploting the energy values on the plane of the potential parameters ($\gamma$--$\rho$--$\nu$), as depicted in fig 4 where the dots represent the number of solutions with real energies, for different values of the parameters, one can easily see the multiplicity and distribution in the energy eigenvalues (and the corresponding wavefunction) for any given eigenstate.

 To our knowledge, such behaviour has not been pointed out before within the context of (quasi-) exact solvablity of quantum systems. Although such behaviour may appear rather uncommon within the context of QES systems, however, we note that this is merely a generalization of \cite{FS2009}, where the authors showed that the complete spectrum for the $\mathcal{H}_2$ model is only obtainable for specific values of the potential parameters ($\omega=1/2$ and $g=2$). Explicitly, we have demonstrated that the eigenstates for higher members of the family have multiple QES sectors which are parameter dependent. Moreover, one other interesting characteristic of the solution is that for some points ($\gamma, n, \ell$), take for instance (1, 1, 1), the solutions space becomes entirely complex (in terms of the energy and the allowed potential parameters). Thus, at such points, the Hamiltonian cease to be hermitian but still quasi-exactly solvable.

\begin{figure}[ht]
\centering
\subfloat[$n=1$, $\ell=1$\label{subfig-1:dummy}]{%
\includegraphics[scale=.6]{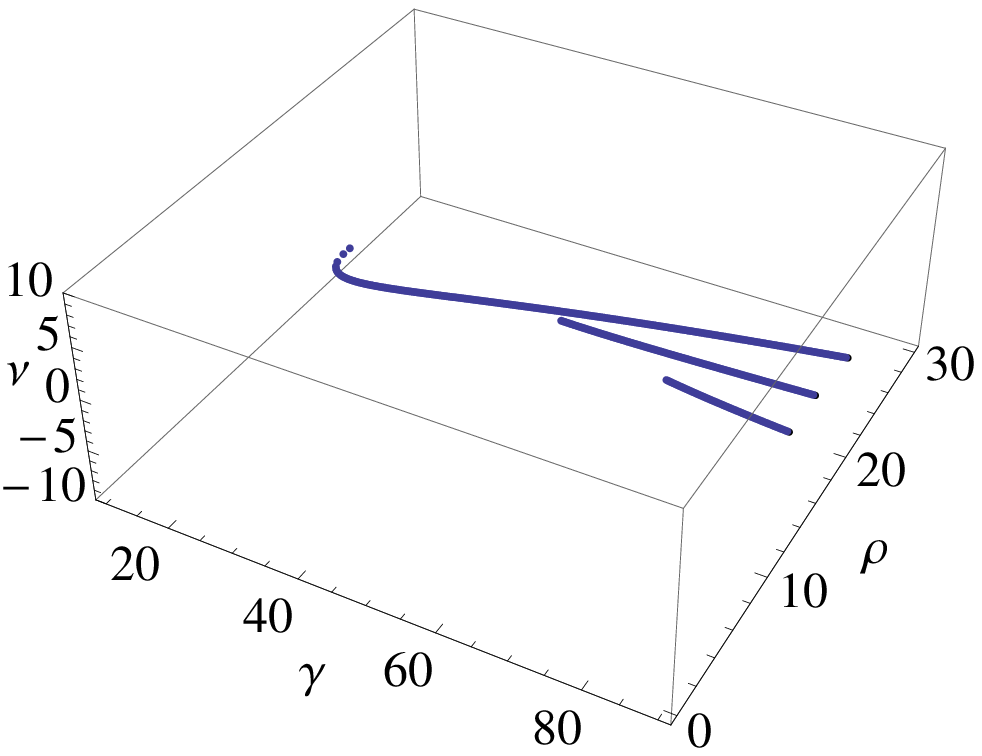}}\hfil
 \subfloat[$n=1$, $\ell=2$\label{subfig-2:dummy}]{%
\includegraphics[scale=.6]{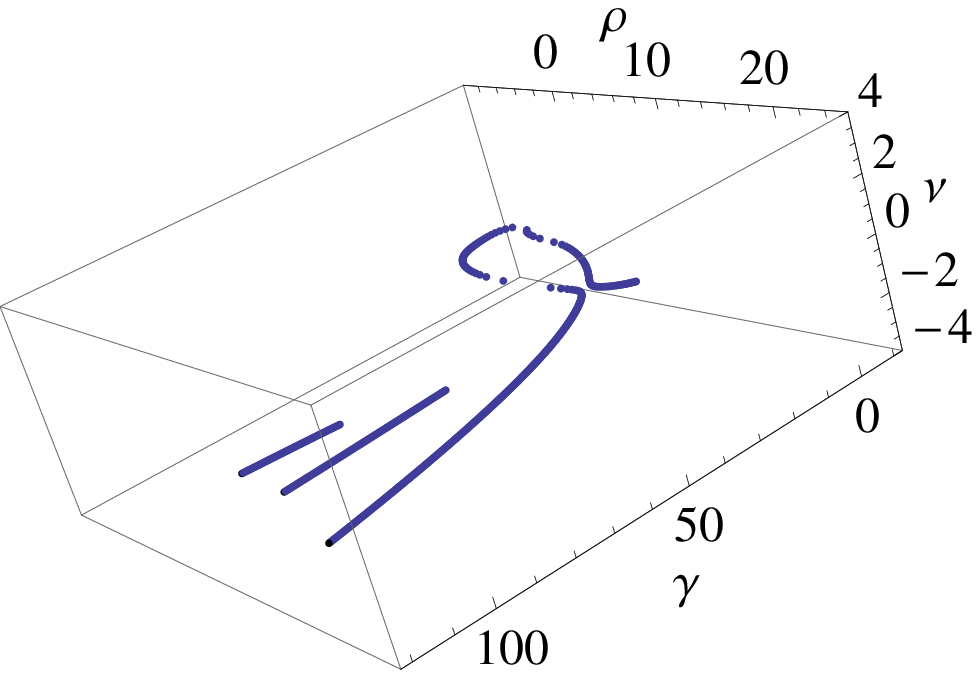}}\\
\subfloat[$n=1$, $\ell=3$\label{subfig-1:dummy}]{%
\includegraphics[scale=.6]{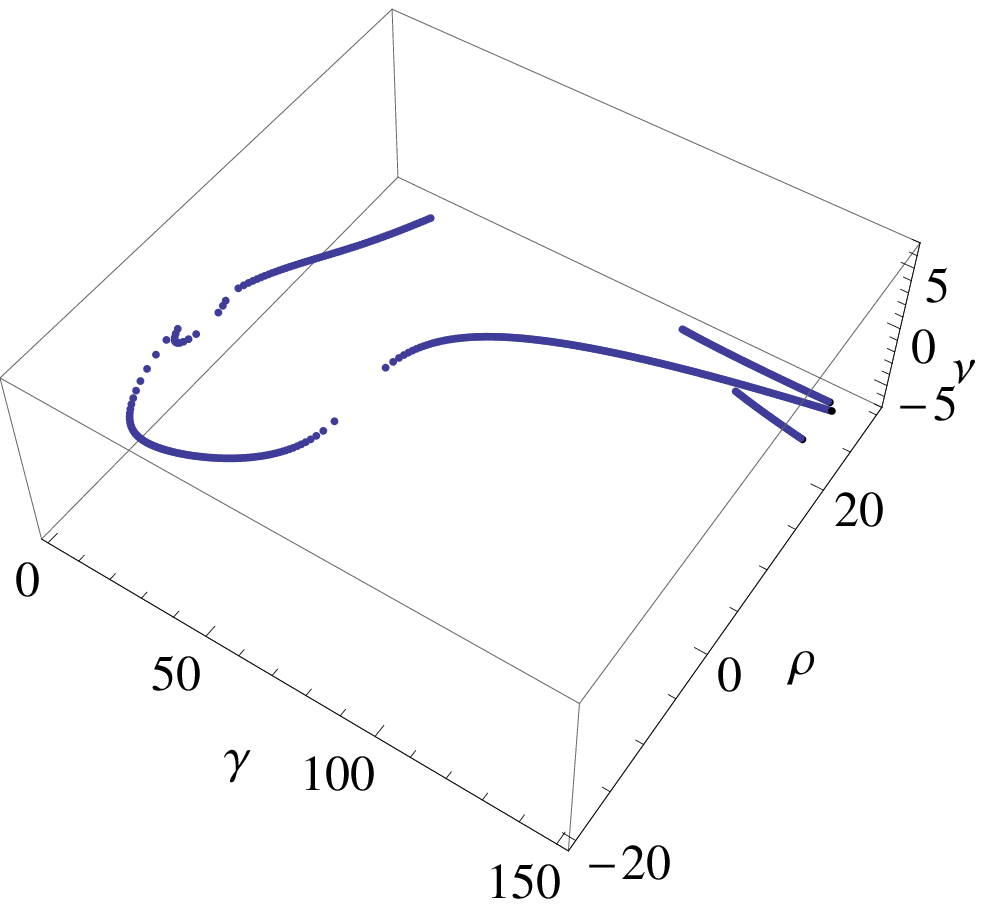}}\hfil
\subfloat[$n=1$, $\ell=4$\label{subfig-1:dummy}]{%
\includegraphics[scale=.6]{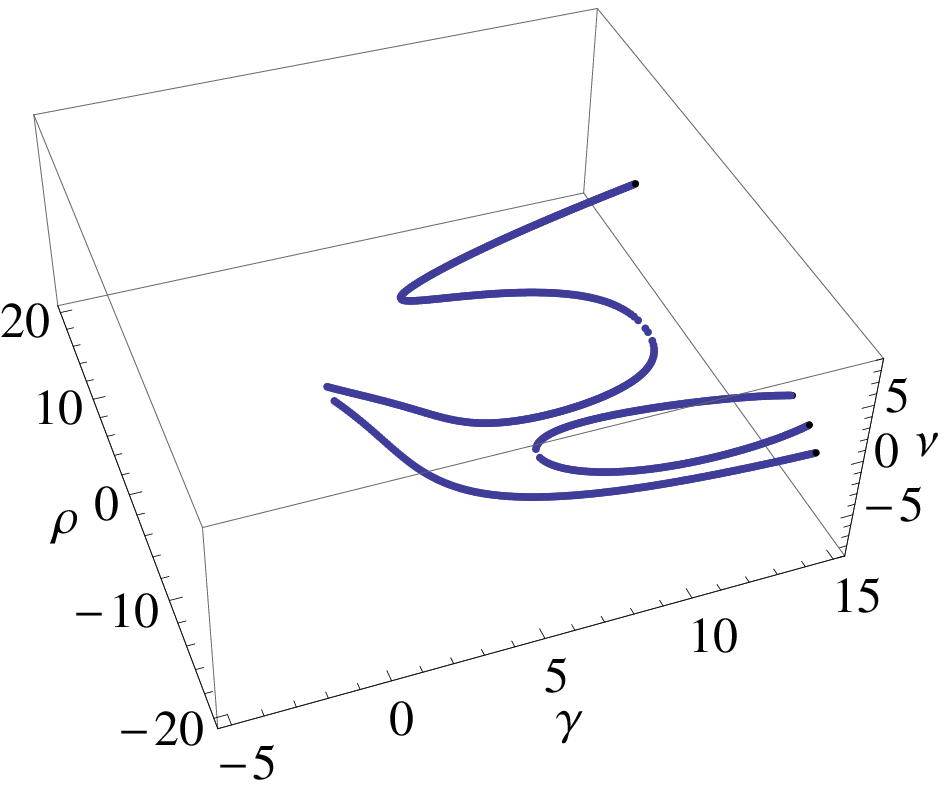}}
\caption{Energy plots for $\mathcal{H}_4$ model for $n=1$ and $1\leq\gamma\leq 100$. These four figures show the variation in the number of QES sectors for each states, with respect to the change in the values of the potential parameters.}
\label{fig:}
\end{figure}
Finally for $n=2$, we have the energy and wavefucntion
\begin{equation}\label{eq:35}
E_2=2\sqrt{\gamma}(z_1+z_2)+3\sqrt{\gamma}\left(2\nu+\ell+\frac{13}{2}\right)+\frac{\rho}{\sqrt{\gamma}}\left(2\nu+\frac{\ell}{2}+\frac{11}{4}\right)+\frac{3\rho^2}{8\gamma}+\frac{3\kappa}{2}$$$$
\Psi_1(r)\sim r^{\ell+1}(4r^4+12r^2+3)^\nu e^{-\frac{\sqrt{\gamma}}{4}r^4-\frac{\rho}{4\sqrt{\gamma}}r^2}(r^2-z_1)(r^2-z_2)
\end{equation}subject to the constriants
\begin{equation}\label{eq:36}
\kappa=\frac{\rho^2}{4\gamma}-\sqrt{\gamma}(2\ell+8\nu+13)$$$$
-2\sqrt{\gamma}(z_1^3+z_2^3)-\left(6\sqrt{\gamma}+\frac{\rho}{\sqrt{\gamma}}\right)(z_1^2+z_2^2)+\left(2\ell+8\nu-\frac{3\rho}{\sqrt{\gamma}}-\frac{3\sqrt{\gamma}}{2}+7\right)(z_1+z_2)$$$$
+2\left(6\ell+12\nu-\frac{3\rho}{4\sqrt{\gamma}}+15\right)+\frac{3E_2}{4}+3\nu(2\ell+\nu+2)=0,$$$$
6E_2-12\nu(1-\nu)-\frac{3\kappa}{4}=4\sqrt{\gamma}(z_1^2+z_2^2)+\left(6\sqrt{\gamma}+\frac{\rho}{\sqrt{\gamma}}\right) (z_1+z_2)-2(4\ell+16\nu-3\sqrt{\gamma}+10)$$$$
+\frac{\rho}{\sqrt{\gamma}}\left(12+6\nu+2\ell-\frac{3\rho^2}{16}\right)-(2\ell+5)\left(4\nu-\frac{3\sqrt{\gamma}}{4}\right),
\end{equation}
with the roots $z_1,z_2$ satisfying the equations
\begin{equation}\label{eq:37}
\footnotesize\sum_{j\neq i}^2\frac{2}{z_i-z_j}=\frac{-4\sqrt{\gamma}z^4_i-\left(12\sqrt{\gamma}+\frac{2\rho}{\sqrt{\gamma}}\right)z^3_i+\left(6-3\sqrt{\gamma}+4\ell+16\nu-\frac{6\rho}{\sqrt{\gamma}}\right)z^2_i+3\left(6+4\ell+8\nu-\frac{\rho}{2\sqrt{\gamma}}\right)z_i+3\left(\ell+\frac{3}{2}\right)}{z_i(4z_i^2+12z_i+3)},~i=1,2.
\end{equation}

\section{Conclusions}
In summary, we have discussed the bound-state solutions to a family of isotonic oscillators is based on the pseudo-Hermite EOPs. We showed that the corresponding Schr\"odinger equation for the first two members of the family is reducible to QES differential equations. Using the Bethe ansatz approach, we systematically obtained  the exact closed-form  energies, wavefunctions and allowed potential parameters for these member oscillators. 

We pointed out some interesting properties exhibited by these oscillators. 
In addition, extensive numerical computations reveal that member potentials have multiple quasi-exactly solvable eigenstates and the number of states for higher members are parameter dependent. It is pertinent to note that though our method gives general quasi-exact solutions for these models for all allowed valueds of the potential parameters, however, not all of them yield a physical solutions. 

It would be interesting to extend the present work to the QES models which are based on type I, II or III Laguerre EOPs and two-step extensions of harmonic oscillator related to $X_{m_1,m_2}$ Hermite EOPs. Research along this path is underway, and the results will be reported elsewhere.


\section*{Acknowledgements}
DA acknowledges the support of the Australian
IPRS and a University of Queensland Centennial Scholarship. The research of IM was supported by the Australian Research Council through DECRA project E130101067. JL and YZZ are supported in part by the Australian Research Council through Discovery Project DP110101414 and DP110103434 respectively.

\begin{appendix}
\section{Generalized Bethe ansatz method}
Here we give a description of the Bethe ansatz method \cite{YZZ2012}. We consider the second order differential equation
\begin{equation}\label{eq:gbm1}
\left[P(z)\frac{d^2}{dz^2}+Q(z)\frac{d}{dz}+W(z)\right]S(z)=0,
\end{equation}
where $P(z), Q(z)$ and $W(z)$ are polynomials of degree ($t<s$)
\begin{equation}\label{eq:gbm2}
P(z)=\sum_{k=0}^rp_kz^k,\hspace{0.2in}Q(z)=\sum_{k=0}^sq_kz^k,\hspace{0.2in}W(z)=\sum_{k=0}^tw_kz^k,
\end{equation}
$p_k, q_k$ and $w_k$ are constants. If we seek a polynomial solution of the form
\begin{equation}\label{eq:gbm3}
S(z)=\prod_{i=1}^n(z-z_i),\hspace{0.3in}S(z)\equiv 1\hspace{0.1in}\mbox{for}\hspace{0.1in} n=0,
\end{equation}
then  Eq.\,\eqref{eq:gbm1} becomes
\begin{equation}\label{eq:gbm4}
\sum_{k=0}^rp_kz^k\sum_{i=1}^n\frac{1}{z_i-z_j}\sum_{j\neq i}^n\frac{2}{z_i-z_j}+\sum_{k=0}^sq_kz^k\sum_{i=1}^n\frac{1}{z_i-z_j}+\sum_{k=1}^tw_kz^k=-w_0
\end{equation}
where $\{z_i\}$ are distinct roots of the polynomial solution. 
The right hand side of this equation is a constant, while the left hand side is a meromorphic 
function with simple poles $z=z_i$ and singularity at $z=\infty$. The residue at the simple pole $z=z_i$ are given as 
\begin{equation}\label{eq:gbm5}
\mbox{Res}(-w_0)_{z=z_i}=\sum_{k=0}^rp_kz^k_i\sum_{j\neq i}^n\frac{2}{z_i-z_j}+\sum_{k=0}^sq_kz^k_i,
\end{equation}
such that
\begin{equation}\label{eq:gbm6}
\sum_{k=0}^rp_k\sum_{i=1}^n\left(\frac{z^k-z_i^k}{z-z_i}\right)\sum_{j\neq i}^n\frac{2}{z_i-z_j}+\sum_{k=0}^sq_k\sum_{i=1}^n\left(\frac{z^k-z_i^k}{z-z_i}\right)+\sum_{k=1}^tw_kz^k=-w_0-\sum_{i=1}^n\frac{\mbox{Res}(-w_0)_{z=z_i}}{z-z_i}.
\end{equation}
If we define for $\nu\in\mathbb{Z}_+$,
\begin{equation}\label{eq:gbm7}
\mathcal{M}^\nu[z,z_i]=\frac{z^\nu-z_i^\nu}{z-z_i}=z^{\nu-1}+z^{\nu-2}z_i+\dots+zz_i^{\nu-2}+z_i^{k-1}
\end{equation}and 
\begin{equation}\label{eq:gbm8}
\mathcal{S}[z_i^\nu;z_j]=\sum_{i=1}^n\sum_{j\neq i}^n\frac{z_i^\nu}{z_i-z_j}=\sum_{i=1}^n~\sum_{j=i+1}^nz_i^{\nu-1}+z_i^{\nu-2}z_j+\dots+z_iz^{\nu-2}_j+z_j^{\nu-1}
\end{equation} then Eq.\,\eqref{eq:gbm6} reduces to
\begin{equation}\label{eq:gbm9}
2\sum_{k=1}^rp_k\mathcal{S}\left[\mathcal{M}^k[z,z_i];z_j\right]+\sum_{k=1}^sq_k\sum_{i=1}^n\mathcal{M}^k[z,z_i]+\sum_{k=1}^tw_kz^k=-w_0-\sum_{i=1}^n\frac{\mbox{Res}(-w_0)_{z=z_i}}{z-z_i}.
\end{equation}
For this equation to be valid, the right hand side must also be a constant. By Liouville's theorem, we demand that the coefficients of the powers of 
$z$ as well as the residues at the simple poles of the right hand side be zero. As a result, one can we evaluate the three terms on the l.h.s of  Eq.\,\eqref{eq:gbm9} for some $k$, such that the sums of all possible coefficients of $z^k$ are equated to zero,
\begin{equation}\label{eq:gbm10}
2p_k\mathcal{S}\left[\mathcal{M}^k[z,z_i],z_j\right]+q_k\sum_{i=1}^n\mathcal{M}^k[z,z_i]+w_kz^k+w_0=0,
\end{equation}where $t=1,\dots,s-1$ (necessary for quasi-exact solutions), and the roots $\{z_i\}$ satisfy the Bethe ansatz equations
\begin{equation}\label{eq:gbm11}
\sum_{k=0}^rp_kz^k_i\sum_{j\neq i}^n\frac{2}{z_i-z_j}+\sum_{k=0}^sq_kz^k_i=0.
\end{equation}
Finally, we note that if $P(z)$, $Q(z)$ and $W(z)$ are of degrees 4, 3 and 2 respectively, then the above proceduce reduces to that discussed in ref \cite{YZZ2012}.
\end{appendix}


\begin{thebibliography}{99}
\bibitem{CQ2008} C. Quesne, J. Phys. A.: Math. Theor. {\bf 41} (2008) 392001
\bibitem {OS2009} S. Odake and R. Sasaki, Phys. Lett. B {\bf 679}  (2009) 414.

\bibitem {GKM2010b} D. G´omez-Ullate, N. Kamran, and R. Milson, J. Phys. A: Math. Theor. {\bf 43} (2010) 434016 .
\bibitem {STZ2010} R. Sasaki, S. Tsujimoto, and A. Zhedanov, J. Phys. A: Math. Theor. {\bf 43} (2010) 315204.
\bibitem {GKM2012} D. G´omez-Ullate, N. Kamran, and R. Milson, J. Math. Anal. Appl. {\bf 387} (2012) 410.

\bibitem{MQ2013a}I. Marquette and C. Quesne, J. Math. Phys. 54 (2013) 042102
 
\bibitem{MQ2013b}I. Marquette and C. Quesne, J. Phys. A: Math. Theor. {\bf 46} (2013) 155201 
 
\bibitem{MQ2013c}I. Marquette and C. Quesne, J. Math. Phys. {\bf 54} (2013) 042102 
\bibitem{ODSA2013} S. Odake and R. Sasaki, J. Phys. A: Math. Theor. {\bf 46} (2013) 245201
\bibitem{GGM2014} D. Gomez-Ullate, Y. Grandati and R. Milson, J. Phys. A : Math. Theor. {\bf 47} (2014) 015203
\bibitem{PTV2012} S. Post, S. Tsujimoto, and L. Vinet, J. Phys. A: Math. Theor.
{\bf 45} (2012) 405202.
\bibitem{Turbiner96} A. Turbiner, CRC Handbook of Lie Group Analysis of Differential Equations, Vol. {\bf 3}, Chap. 12, ed.
  N. H. Ibragimov, CRC Press, Boca Raton, FL, 1996.
\bibitem{GKO93} A. Gonz\'arez-L\'opez, N. Kamran and P. Oliver, Commun. Math. Phys. {\bf 153} (1993) 117.
\bibitem{Ushveridze94} A.G. Ushveridze, Quasi-exactly solvable models in quantum mechanics, Institute of Physics
  Publishing, Bristol, 1994.
\bibitem {YZZ2012} Y.-Z. Zhang,  J. Phys. A: Math. Theor. {\bf 45} (2012) 065206.
\bibitem{AZ11} D. Agboola and Y.-Z. Zhang, J. Math. Phys. {\bf 53} (2012) 042101.
\bibitem {DZ12c}D. Agboola and Y.-Z. Zhang, Ann. Phys. (NY) {\bf 330} (2013) 246.
\bibitem{AZMPL12}D. Agboola and Y.-Z. Zhang, Mod. Phys. Lett. A {\bf 27} (2012) 1250112.
\bibitem{CPRS08} J. F. Cari\~nena, A. M. Perelomov, M. F. Ra\~nada and M. Santander, J. Phys. A: Math. Theor. {\bf 41} (2008) 085301.  
\bibitem {FS2009} J. M. Fellows and R. A. Smith, J. Phys. A: Math. Theor. {\bf 42} (2009) 335303.
\bibitem{S2010} J. Sesma, J. Phys. A: Math Theor. {\bf 43} (2010) 185303.
\bibitem{HSY10} R. L. Hall, N. Saad and O. Yesiltas, J. Phys. A: Math. Theor. {\bf 43} (2010) 465304.
\bibitem{SHCY11} N. Saad, R. L. Hall, H. Cifti and O. Yesilatas, Adv. Math. Phys. {\bf 2011} (2011) 750168.





\end{thebibliography}
\end{document}